\documentclass[12pt]{article}

\usepackage{jinstpub} % for details on the use of the package, please
\usepackage{lineno}
\usepackage{hyperref}
\title{\boldmath Initial Calibration of Large Timing Arrays for the LHC}

\author[a]{Sebastian~White}

\emailAdd{sebastian.white@cern.ch}

\affiliation[a]{ University of Virginia \\Charlottesville, Virginia, USA}

\abstract{ 
	In preparation for HL-LHC operation a number of new detector systems are being constructed with
timing precision on physics objects of $\leq50$ picoseconds. These time stamps will reduce the level
of pileup induced backgrounds in this LHC phase where the number of interactions per crossing will
reach of order 100-200\cite{Shiltsev}.
	In the case of CMS, three new systems have initially to be corrected for the usual amplitude walk
resulting from the effect of variations in signal size on leading edge timing. In these systems the resulting
timing spread (ie walk) ranges from one to four nanoseconds.
	In the following note we advocate approaching this initial calibration for walk as a calculable correction
given early calibration during commissioning- rather than depending on special collider data to perform the calibration.
	We derive a simple analytic expression for the walk correction and confirm its effectiveness with lab data. }
	

\begin{document}
\maketitle
\flushbottom
\section{Introduction}
\label{sec:Introduction}

  	Both ATLAS and CMS are preparing detector systems capable of time measurement of $\leq50$ picoseconds precision. In the case of CMS\cite{TDR} the MIP timing detector consisting of
a Barrel Timing Layer(BTL) and and Endcap Timing Layer (ETL) comprise several times $10^5$ channels with a primary role of mitigating pileup induced backgrounds.

	Whereas in earlier collider timing detectors an approach of time calibrating the channels using actual collider data was practical we argue that this would be far more difficult for 
these systems. In CDF\cite{CDF}, for example, it was possible to select a clean data set (absent pileup) with Z->ee to tune the calibration (ie amplitude walk and time offsets) and the total channel
count was of order $10^3$.

	By contrast, in the high pileup environment of the LHC it will likely be difficult to obtain such a clean sample. Furthermore the impact of radiation on detector performance will
be significant resulting in changing operating configuration and therefore re-calibrations. In the case of the BTL (which we focus on in the following discussion) high radiation doses
to the SiPMs and frequent annealing during pauses in collider operation will likely result in changing calibration parameters.

	Fortunately, a major piece of the timing calibration (amplitude walk) lends itself to a simple analytic parametrization which we derive below. Given the right measurement
a simple stand-alone calibration data set could put in place compensation for walk. One system where the needed measurements are clearly in place is the BTL for which an
ASIC capable of providing two leading edge discriminator times and a measure of signal amplitude(referred to as "Q" in the following)\cite{tahereh}. The role of dual thresholds on the leading edge is to provide both time and slope ($\frac{dV}{dt}$).
In practice the slope measurement needs only be used in limited calibration runs to establish an average correspondence to Q.

\begin{figure}[!htp]
%%\centering
\includegraphics[width=.95\textwidth]{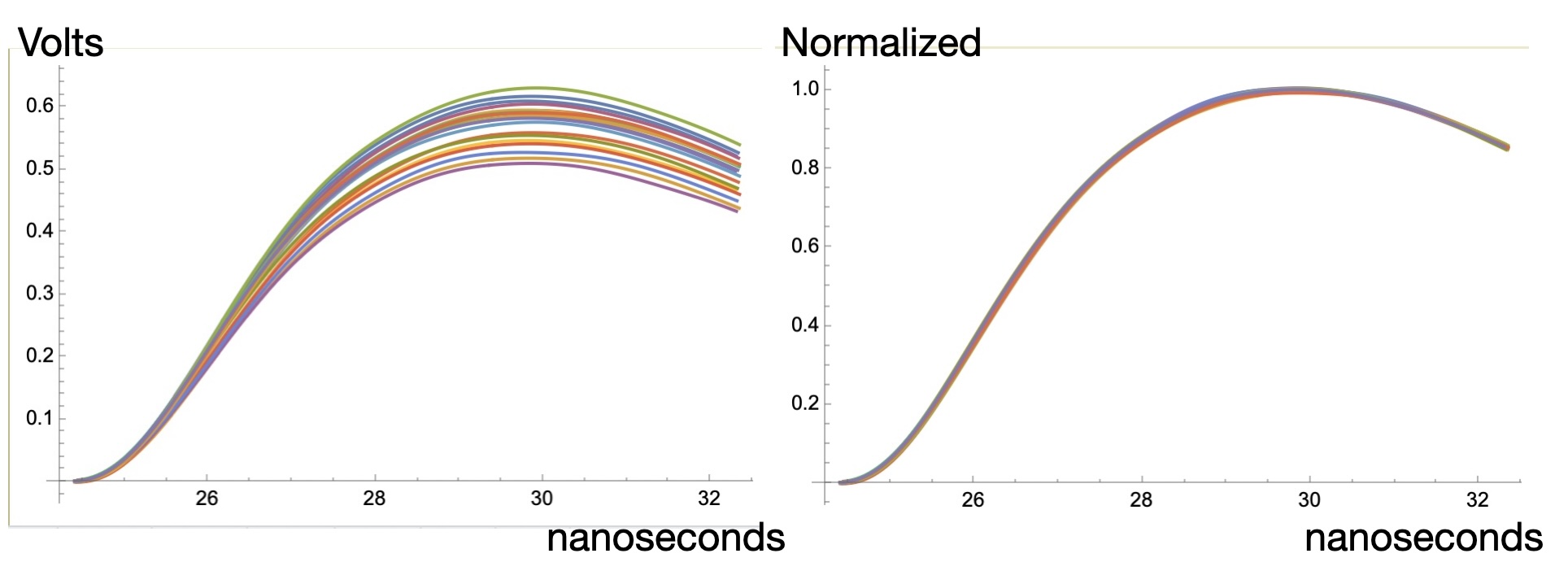}
\caption{The principle of the Constant Fraction technique on captured waveforms. Several events displayed in the left panel are rescaled according to pulse area resulting in a consistent 
constant fraction time. In the following discussion we choose a particular timing threshold of 0.2 Volts.}
\label{fig:wavefor}       
\end{figure}

\section{Ideal Constant Fraction Timing}
\label{sec:CF}

	The issue of Amplitude walk\cite{Knoll} is finessed when it is possible to renormalize all signals to the same amplitude and set a timing threshold
corresponding to a fixed fraction of the signal peak value (see Figure 1). Since data acquisition tools to capture the full waveform (including digital scopes) are
commonly available and often used in performance tests of new timing detectors we have a lot of experience with this Constant Fraction technique.
In this paper we will use laboratory data originally intended to evaluate signal processing  of SiPMs intended for the BTL. 
	
	We have used this Constant Fraction technique on a variety of timing detectors including BTL LYSO/SiPM assemblies, High Gain Silicon and 
the MPGD based PICOSEC project. In all cases this technique achieves ultimate time resolution. Therefore we base the remainder of this report on that technique.

	As we will see in the following, we can accomplish this analysis without the full waveform. Instead with a very limited amount of information we can
derive the walk correction as if we were performing constant fraction analysis.

\section{Data Set Used for this Paper}

	For most of the discussion below we refer to a set of laboratory measurements reported earlier \cite{IPRD} employing a 3x3 mm$^2$ SiPM\footnote[1]{Hamamatsu  HPK S12572 - 015 SiPM} with a high bandwidth transimpedance amplifier. Waveforms were recorded using a 1 GHz analog bandwidth, 20 GSa/s digital oscilloscope. The SiPM was illuminated from a PicoQuant Laser head with 470 nm peak emission and a pulse width of 350 picoseconds (rms).
	
		Finally, the "time reference" which is used to confirm the successful walk correction is derived from the laser trigger.
	
\begin{figure}
\centering
\centerline{\includegraphics[width=0.75\textwidth]{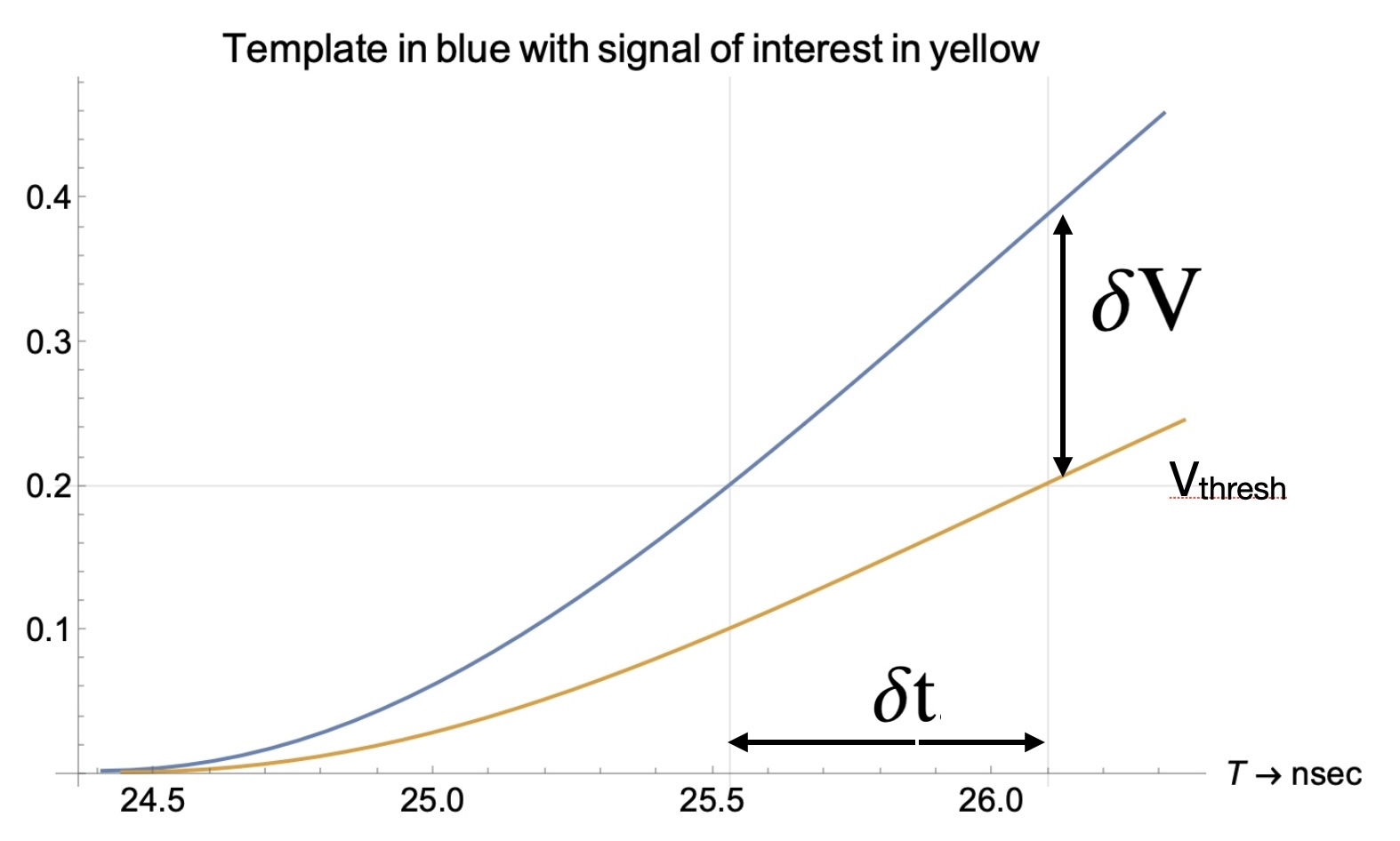}}
\caption{Derivation of the analytic form for walk correction from the Constant Fraction model. In this particular event the signal crosses the timing threshold (V$_{thresh}$) at t=26.1 nanosec. We will calculate the required time shift $\delta$t, to obtain the Constant Fraction time on the template signal. Since $\delta$V is related to V$_{thresh}$ by the scaling $\frac{dV+V_{thresh}}{V_{thresh}}$=$\frac{Q_{template}}{Q}$ the time offset is obtained from $\frac{dV}{dt}$=the template slope.}
\label{fig:modelmodeln}       
\end{figure}

\section{Analytic expression for the time walk correction}

	Since, to be specific, we are using terminology corresponding to the BTL ASIC (TOFHIR2) we list below the data elements generated for each event:
\begin{itemize}
\item{the time stamp for leading edge discriminator T1- the raw signal time to be corrected for amplitude walk}
\item{ the time stamp for a second leading edge discriminator with a higher threshold setting- to be used in determining slope for calibration data}
\item{ a measurement of signal amplitude derived from integrating a clipped version of the signal called "Q" and known to be reliable to 3$\%$}
\end{itemize}

	Referring to Figure 2 it is apparent that much less information than the full waveform is needed to accomplish the walk correction.
	For the case of the BTL, where the slope of pulses at threshold can be measured, a small stand alone data set can provide the needed $\frac{dV}{dt}$ of the template slope.
Evidently	
	
\begin{equation}
\frac{dV_{template}}{dt}=<\frac{dV_i}{dt}/Q_i>\times Q_{template}
\end{equation}

	where the brackets <> are meant to denote a sample mean of, say, $10^3$ events where we record the slope ($\frac{dV_i}{dt}$ ) determined from the times of two leading edge threshold
discriminators as well as the charge ($Q_i$) output of the BTL ASIC.
	
	In the case of BTL we would simply collect a data set to establish a mean ratio of slope to charge (Q). Then we choose a template with a particular $Q_{template}$ and use that and its derived slope to obtain event-by-event walk corrected times for a given channel.

\begin{equation}
\delta t=(1-\frac{Q_{template}}{Q})\times V_{thresh} \times (\frac{dV_{template}}{dt})^{-1}
\end{equation}

	It should be emphasized that the time walk correction is obtained from a single measurement for any given event (Q) once the template scale ($Q_{template}$) has been fixed and
the corresponding slope ($\frac{dV_{template}}{dt}$) has been calibrated. The timing discriminator T1 setting ($V_{thresh}$) for each channel is controlled by a programmable DAC and presumably known.
	
	Clearly this expression gives a simple prescription  for amplitude walk, We next test it using the sample of lab data. The following plots (see Figures 3 and 4) illustrate the effectiveness of the
technique. We also note that the final time spread gives the expected value 20 picosecond rms.

\begin{figure}[!htp]
%%\centering
\includegraphics[width=.45\textwidth]{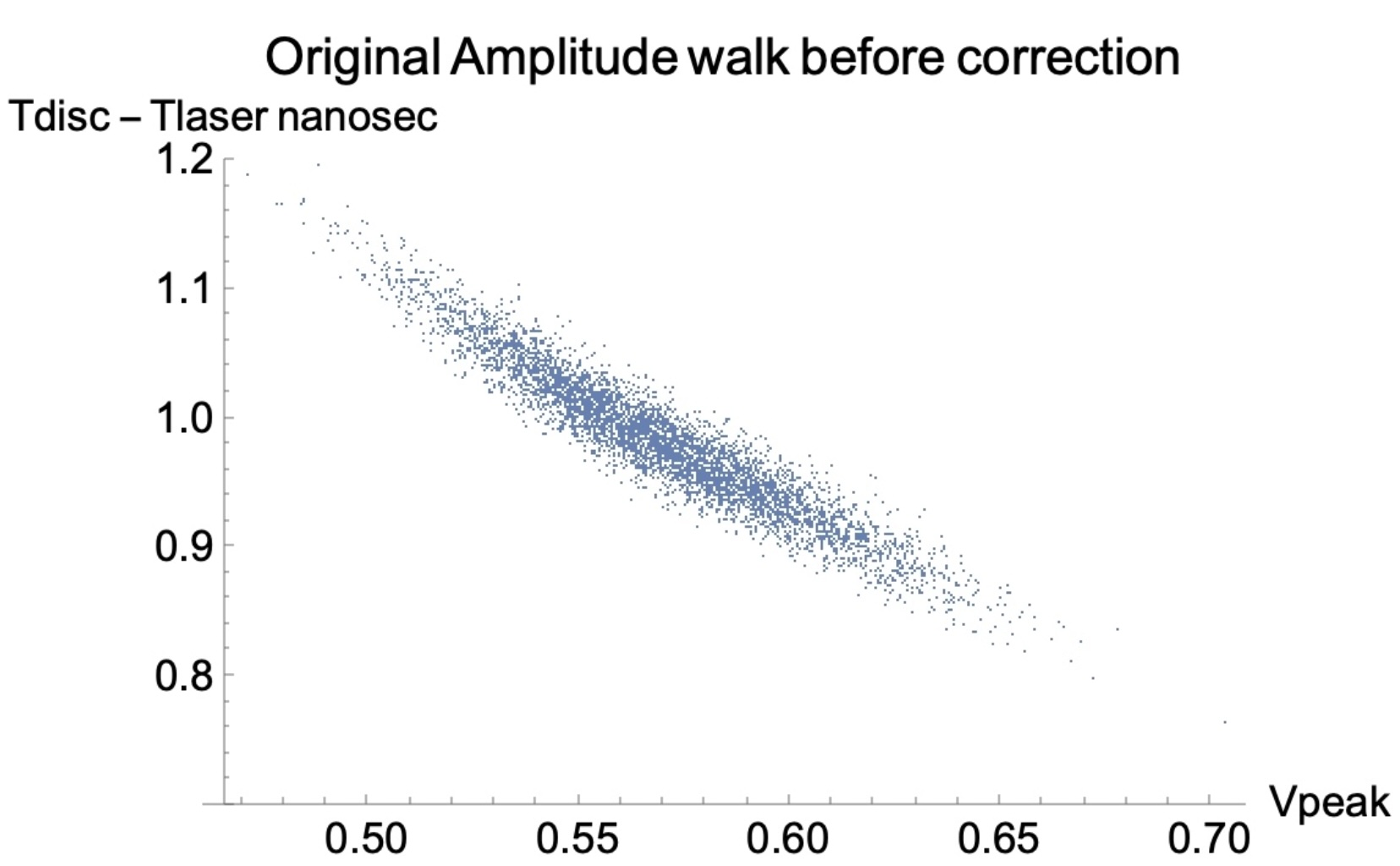}
\includegraphics[width=.45\textwidth]{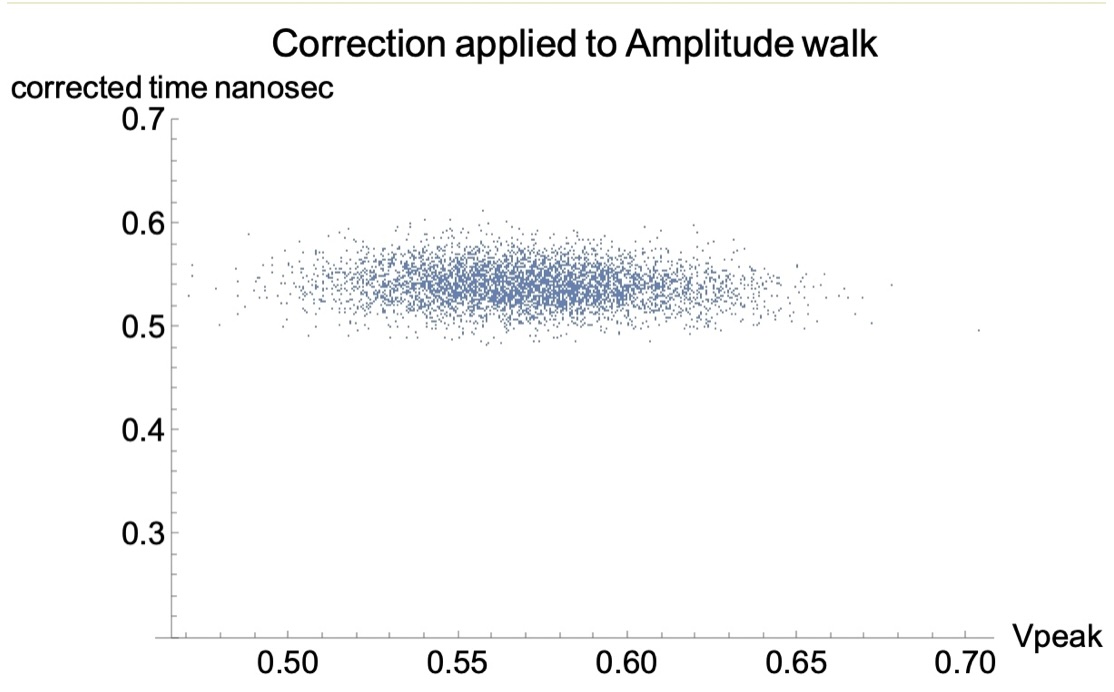}
\caption{Returning to the lab data we apply eqn. 4.2 to correct events for amplitude walk and verify the correction by referring to the laser trigger as a t0.}
\label{fig:wavefor}       
\end{figure}

\begin{figure}[!htp]
%%\centering
\includegraphics[width=.85\textwidth]{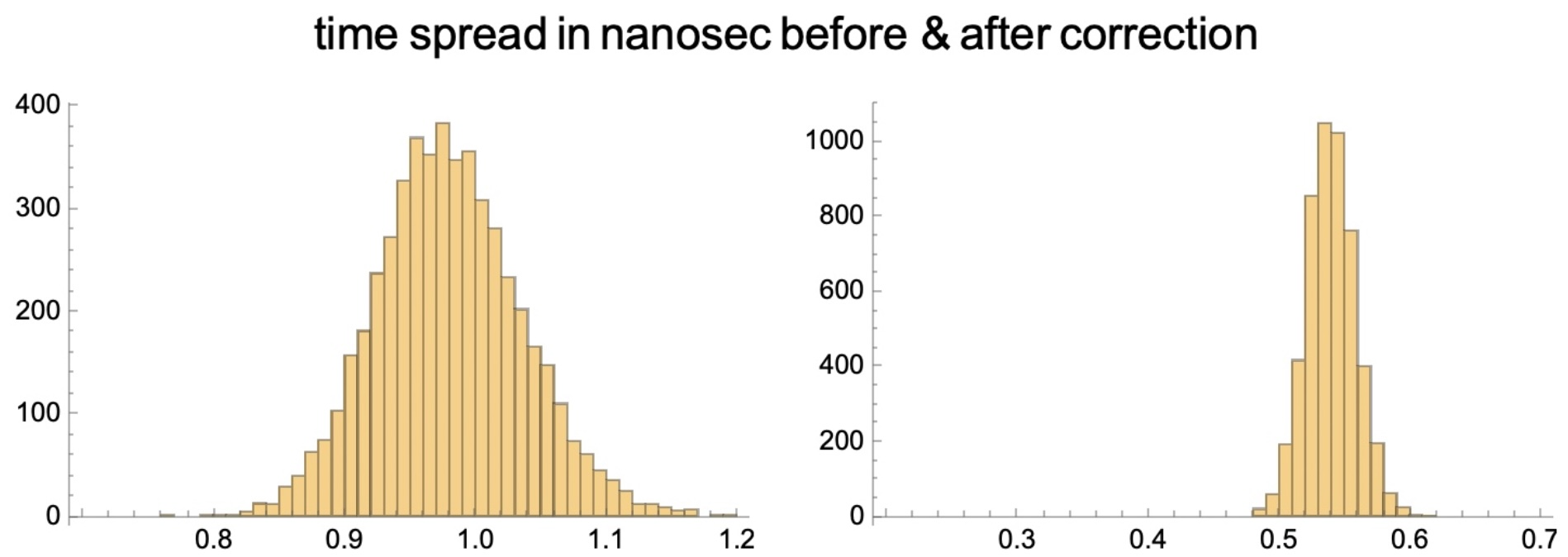}
\caption{The previous figure represented as a histogram of the corresponding time spreads- in nanoseconds. Note that for the corrected distribution the rms time spread is 20 picoseconds which is
expected to be the ultimate resolution.}
\label{fig:wavefor}       
\end{figure}
	
\section{Conclusion}

	We have demonstrated a simple analytic procedure for correcting the large time spread of signals due to amplitude walk. In the case of the CMS BTL system the relevant
calibration parameters can readily be determined in small stand alone data sets without reference to other detectors. It is likely that other systems have tools for deriving the 
template slope parameter. We strongly encourage pursuing this approach since the complexity of events at HL-LHC, the large number of timing channels involved and the
likely need for occasional re-calibrations certainly make the possible alternative of an analytical prescription for the walk correction attractive.

This work received partial support through the US CMS program under DOE contract No. DE-AC02-07CH11359.

\end{document}